\documentstyle[psfig,sprocl]{article}

\def\Journal#1#2#3#4{{#1} {\bf #2}, #3 (#4)}

\def\PRL{\em Phys. Rev. Lett.}
\def\PRA{{\em Phys. Rev.} A}
\def\JPB{\em J. Phys. B: At. Mol. Opt. Phys.}
\def\JPA{\em J. Phys. A: Math. Gen.}
\def\be{\begin{equation}}
\def\ee{\end{equation}}
\def\bea{\begin{eqnarray}}
\def\eea{\end{eqnarray}}

\def\vereq#1#2{\lower3pt\vbox{\baselineskip1.5pt \lineskip1.5pt
\ialign{$#1\hfill##\hfil$\crcr#2\crcr\sim\crcr}}}

\begin{document}

\renewcommand{\textfraction}{0.2}
\renewcommand{\floatpagefraction}{0.8}

\title{ATOMS IN STATIC FIELDS:  CHAOS OR DIFFRACTION ?}

\author{P.A. \  DANDO and  T.S.\  MONTEIRO}

\address{Department of Physics and Astronomy, University College London,
Gower Street, London WC1E 6BT, United Kingdom}

\maketitle

\abstracts{
A brief review of the manifestations of classical chaos 
observed in atomic systems is presented.  Particular attention is
paid to the analysis of atomic spectra by periodic 
orbit-type theories.  For diamagnetic non-hydrogenic Rydberg atoms,
the dynamical explanation for observed spectral features has been
disputed.  By building on our previous work on the photoabsorption
spectrum, we show how, by the addition of diffractive terms, the
spectral fluctuations in the energy level spectrum of general Rydberg
atoms can be obtained with remarkable precision from the Gutzwiller trace
formula. This provides further evidence that non-hydrogenic systems 
are most naturally described in terms of diffraction rather than 
classical chaos.
}

\section{Atoms as Laboratories of Quantum Chaos}

Atomic physics has provided some of the most important examples of
real, experimentally observable,  quantum systems for which the 
underlying classical motion is chaotic.  
In broad terms, the study of chaos in atoms comprises two major strands.
One is the study of time-dependent periodically driven systems where the
phenomenon of dynamical localization provides a mechanism for the quantum
suppression of chaotic diffusion.  This is exemplified by the behaviour of
hydrogen in a microwave field~\cite{Koch} and, more recently, atoms in
traps.\cite{Raizen}  The other strand is the study of time independent
systems.  Here, highly excited atoms in static external fields have
received particular attention.  Although chaos in the sense of exponential
sensitivity to perturbations is not present, atomic spectra exhibit 
the characteristic `footprints' of classical chaos, such as
eigenvalue statistics similar to those of random matrices, 
spectral modulations and `scarring' of wavefunctions by unstable 
periodic orbits. 

\begin{figure}[htb]
\centerline{
\psfig{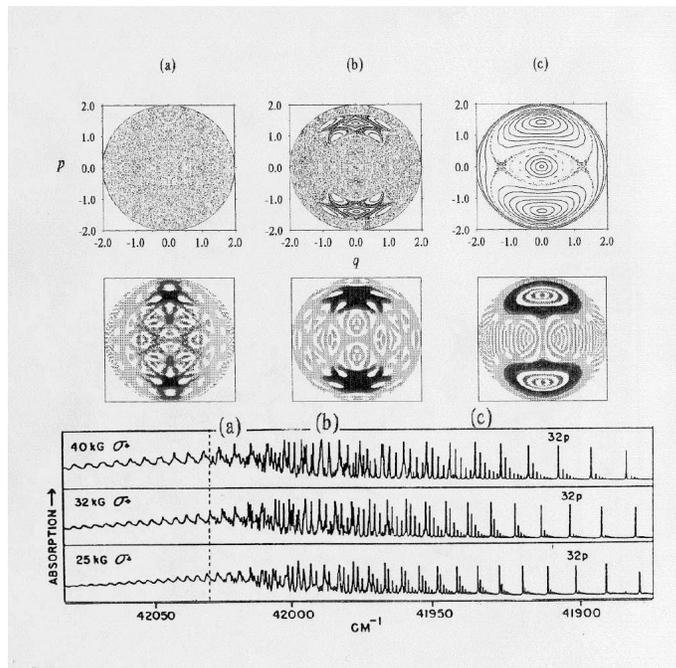}
}
\caption {The classic Garton-Tomkins spectrum of barium in a 
magnetic field.  Above are shown are Wigner
functions and classical Poincar\'{e} surfaces-of-section for
diamagnetic hydrogen at scaled energies  (a)~$\varepsilon=-0.1$,
(b)~$\varepsilon=-0.316$ and (c)~$\varepsilon=-0.5$ showing the
gradual transition to chaos as $\varepsilon$ increases.}
\label{fig.gt}
\end{figure}

The classic 1969 Garton-Tomkins spectrum of barium in a magnetic
field~\cite{GT69} revealed the first `footprints' of
classical orbits in the quantum spectrum of a `real' system.  
Oscillations observed in the $m=1$ spectrum near the ionization limit
at energy spacing $\sim 1.5\hbar\omega$ were, much later, associated with 
the periodic orbit perpendicular to the magnetic field.  In
Fig.~\ref{fig.gt} we show the Garton-Tomkins spectrum.  Above are
shown Wigner functions and classical Poincar\'{e} surfaces-of-section
for diamagnetic hydrogen in three different regimes of energy. The
Wigner functions show three different stages of the Garton-Tomkins
orbit: (a)~scar, (b)~bifurcation and (c)~torus quantization.  The
surfaces-of-section show the gradual transition from regularity to
chaos as the scaled energy increases.

Such is the experimental resolution now obtainable 
(being typically of the order of 50MHz) that atoms still provide one 
of the best `laboratories' for investigating and testing theories of
`quantum chaology', although there is also now great interest in chaos in
mesoscopic devices.

For an atom in a strong external field the typical  spectrum,
involving highly excited states in the classically chaotic regime, 
looks extremely irregular.  Eigenstates generally defy any form of 
classification in terms of quantum numbers.  However, in the
semiclassical limit ($\hbar\rightarrow 0$) periodic orbit theory, in
the guise of the Gutzwiller Trace Formula (GTF),\cite{G90} exposes the
beautiful connection between modulations in the quantum spectrum and
classical orbits. Because of a useful scaling property, 
Fourier transforms of calculated atomic spectra can be related directly to
classical stabilities. Hence the GTF has provided atomic physicists with
an extremely powerful framework for the analysis of quantum spectra in 
the classically chaotic limit.

However it is not widely appreciated that there exist two 
separate semiclassical theories for the analysis of atomic spectra.  
The GTF relates classical periodic orbits to
oscillations in the {\em energy level\/} spectrum in the
semiclassical regime.  {\em Closed orbit\/}
theory,\cite{DuD88b,GD92b} on the other hand, is the
appropriate tool for the investigation of  experimental 
{\em photoabsorption\/} spectra (which involve a quantum spectrum
weighted by the appropriate absorption probability).  
\begin{figure}[htb]
\centerline{
\psfig{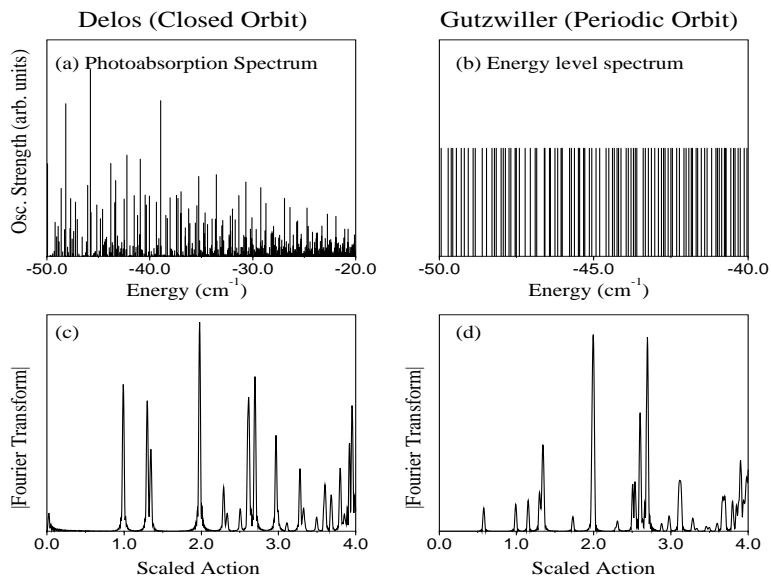}
}
\caption {Comparison between calculated $m=0$, even parity 
(a)~photoabsorption and (b)~energy level spectra for diamagnetic hydrogen.  
(c)~and (d)~show the corresponding 
Fourier transforms which reveal the periodicities in the 
spectral modulations.  Note in particular the differences in the heights
of the peaks and the additional modulations seen in the energy level spectrum
case.}
\label{Fig1a}
\end{figure}

There are
important differences between the two theories which  are
illustrated qualitatively in Fig.~\ref{Fig1a}. These become especially
significant when we seek to understand the differences in behaviour between
hydrogen and other atoms. These differences in behaviour have been
given different dynamical interpretations by different authors. We
review first the main arguments concerning non-hydrogenic spectra,
before contrasting closed orbit theory with periodic orbit theory.

\section{Chaos Versus Diffraction in Non-Hydrogenic Atoms}

It is now 10 years since it was first shown that the GTF provides a
quantitative description 
of highly excited hydrogen atoms in strong external fields.\cite{W87}
However, comparisons between accurate quantal 
spectra reveal spectral amplitudes for non-hydrogenic atoms that
differ substantially from those of hydrogen together with additional
modulations that cannot be associated with any hydrogenic 
orbits.\cite{MW90,Jans93}
Hence, it has been a long-standing problem to apply the GTF
to other species of singly excited (Rydberg) atoms.

Theoretical calculations~\cite{Jans93} found the differences
between hydrogen and other diamagnetic atoms to be  most marked in
the low energy regime where hydrogen is regular. Here, non-hydrogenic atoms
show many `footprints' of chaos.  For example, spectral statistics
lie close to the Wigner (chaotic) 
limit,\footnote{This has been recently verified by
experiment.\protect{\cite{Held97}}} suggesting underlying
chaotic classical motion. However, quantum phase-space
distributions (Wigner functions)
do not have the more `ergodic' appearance 
of those of hydrogen in the classically chaotic regime.
Instead, torus-like structures are found
but with each eigenstate now linked to several `tori' rather than
just one as in the hydrogenic case;  clearly this is not the
signature of underlying classical chaos.

In 1994, the additional spectral modulations (`core-scattered' peaks) 
seen for non-hydrogenic atoms were in fact observed 
experimentally by three separate
groups~\cite{C94,DD94,R94} and identified~\cite{DD94} as arising from
combinations of hydrogenic orbits.

Some progress towards the goal of adapting the GTF to describe
all atoms was made in 1995 when we successfully
adapted the related closed orbit theory of Delos and
co-workers,\cite{DuD88b,GD92b} which describes the
experimentally measured {\em photoabsorption\/} spectrum, to general
non-hydrogenic atoms.\cite{Dando95,Dando96}  However, this work gave
only limited insight and little qualitative explanation of the role played
by the non-hydrogenic core in the actual {\em energy level\/} spectrum.

Classical calculations using a model
potential to describe the effect of the ionic
core~\cite{C94,Main95,Main96} suggest one possible 
explanation, namely that the presence of
the core drastically destabilizes the classical motion.  This
gives rise not only to a reduction in amplitude of the oscillations
associated with classical hydrogenic orbits but also to a multitude of
new orbits not found in the hydrogenic problem.  Although this
approach provides physical insight into the effect of the ionic
core, quantitative agreement is generally poor, 
even failing  to reproduce the quantum results for smaller atoms
such as helium.\cite{pecam}  Indeed, any sort of classical description
of the core-scattering process may be questionable for atoms such
as helium or lithium, where only one 
or two partial waves are influenced by
the core. 

Our 1995 work suggested an alternative explanation of these
phenomena.  In non-hydrogenic atoms both the size of the core
and the characteristic de Broglie wavelength are of the order of
1~au.  Thus it is more natural to describe core effects 
in terms of a {\em diffractive\/} process rather than
classical paths and their stability parameters.
(Recall that diffraction occurs when an obstacle is encountered whose
size is smaller than the wavelength of the incident wave.)  

There is now new evidence that this is indeed the case.  
Now,\cite{Dando97} we have shown that spectral fluctuations
of general Rydberg 
atoms are given with remarkable precision (to within $1\%$)
on including diffractive 
corrections to the GTF.  Several previously unknown features in the
non-hydrogenic energy level spectra were found.  There are additional
modulations that are neither periodic orbits nor combinations of 
periodic orbits.  Also, while `core-shadowing' generally 
decreases the amplitude of oscillations associated with primitive
periodic orbits it can also lead to increases.  In addition, the
spectral statistics of diamagnetic non-hydrogenic atoms,  previously
thought to lie at the Wigner limit, have been found~\cite{DD97} to
belong to an entirely new generic distribution
(`Half-Poisson') usually associated with diffractive systems.

\section{Closed Orbit Theory Versus Periodic Orbit Theory}
\subsection{Closed Orbit Theory}

Closed orbit theory has been presented elsewhere in great 
detail~\cite{DuD88b,GD92b} so we
give only an outline.  Briefly, when an atom absorbs a photon,
the electron propagates outwards in a near zero-energy Coulomb
wave. At sufficiently large distances from the nucleus the wave
propagates semiclassically along classical trajectories. Eventually,
the trajectories and their associated waves are turned back by the
action of the external field.  Some of the trajectories return to the
vicinity of the nucleus and the waves associated with them interfere
with the outgoing waves generating oscillations in the absorption
spectrum.   
The result is a formula for the average oscillator-strength density
that can be written as a combination of a smooth background plus
an oscillatory term,
\begin{equation}
f_{\rm osc}(E) = {\rm Im}\, \sum_{n} \sum_{k} C_{k}^{n}(E) 
       \exp [ i(S_{k}^{n}(E)/\hbar -\pi\mu_k^n /2 -3\pi/4)],
\label{eq:coformula}
\end{equation}
which arises from the interference of semiclassical waves associated
with closed orbits of an electron of energy $E$.
Each different orbit is indexed 
by $k$ and the repetitions of each orbit are labelled by $n$.  
The `recurrence amplitude',  $C_{k}^{n}(E)\propto m_{12}^{-1/2}$,
contains information about the stability of the orbit via the $m_{12}$
element of the stability matrix, $M$;  this is obtained by 
studying the classical motion in the neighbourhood of the orbit.
The phase depends on the classical action, 
$S_{k}^{n} = n \oint_k \, \mbox{\boldmath{$p$}} 
\cdot \,d\mbox{\boldmath{$q$}}$, 
plus an additional term that is computed from the Maslov
Index, $\mu_{k}^{n}$, and other geometrical considerations.

Whereas closed orbit theory describes the {\em photoabsorption\/}
spectrum in terms of the properties of classical orbits
that {\em close\/} at the nucleus, the Gutzwiller Trace
Formula~\cite{G90} relates periodicities in the quantum {\em density
of states,\/} $\rho(E) =-(1/\pi) {\rm Im\, Tr}\, G(E)$,
to isolated {\em periodic\/} orbits.
In the semiclassical limit ($\hbar\rightarrow 0$) 
$\rho(E)$ can also be written as the sum of a smooth background term together
with an oscillatory part of the form,
\begin{equation}
\rho_{\rm osc}(E) = -\frac{1}{\pi}{\rm Im\,} \sum_p \frac{T_p(E)}{i\hbar}
\sum_n \frac{\exp[i(S^n_p(E)/\hbar -\mu^n_p\pi/2)]}
{|2 -{\rm Tr}\, M_p^n|^{1/2}},
\label{gtf}
\end{equation}
where $S^n_p$ again denotes the classical action,
$T_p$ the period, and $\mu_p$ the Maslov index 
and $M_p^n$ the stability
matrix for the $n\,$th repetition of the $p\,$th periodic orbit.

Although Eqs~(\ref{eq:coformula}) and~(\ref{gtf}) show that the
periodicity of the spectral modulations is the same for both
theories there are important differences.  Firstly, the sum in
Eq.~(\ref{eq:coformula}) only includes contributions from orbits 
which deliver amplitude back to the vicinity of the well-localized
initial state:  only orbits that return to the nucleus
contribute.  Note that these orbits need not be periodic.
For the GTF, {\em all\/} periodic orbits of the classical system 
contribute to the sum in Eq.~(\ref{gtf}).  Secondly, the amplitudes
of the contributions are different.  Both formulae contain information
about the stability of the orbit via the stability matrix, $M$.
However, for closed orbit theory, the
recurrence amplitude is related to $m_{12}$--- an off-diagonal
element of $M$---while in the GTF, the amplitude of the modulation
due to a given orbit is proportional to $(2 - {\rm Tr}\,M)^{-1/2}$.

\subsection{Gutzwiller Trace Formula with Diffraction}

The periodic orbit theory of diffraction was developed recently for 
Hamiltonians with discontinuities.\cite{Vatt,Whelan} 
For such systems, periodic orbits are decomposed into two types:
those that do not intersect the discontinuity ({\em geometric\/} orbits)
and those that do ({\em diffractive\/} orbits).  The density of
states is then obtained as a sum:
\begin{equation}
\rho(E) = \underbrace{-\frac{1}{\pi} {\rm Im} \,{\rm Tr}\, 
G_g(E)}_{\rm geometric}  - 
\underbrace{\frac{1}{\pi} {\rm Im}\, {\rm Tr}\, 
G_D(E)}_{\rm diffractive}.
\end{equation}
Taking the trace over the first (geometric) term yields the well-known 
GTF. The trace over the second (diffractive) contribution has been 
shown to be~\cite{Vatt,Whelan}
\begin{equation}
{\rm Tr}\,G_D(E)=   \sum_p \frac{T_p}{i\hbar}\prod_n d(n) G(q_n,q_{n+1};E),
\label{diffrac}
\end{equation}
where $T_p$ is the total sum of periods taken over the paths between 
the vertices and $d(n)$ is the diffraction constant which depends on 
the type of diffraction.  Equation~(\ref{diffrac}) 
encapsulates the important result 
that the trace integral taken between the $n$\/th and $n+1$\/th diffractive
points is proportional to the Green's function between those points.

\subsection{Application to Non-Hydrogenic Atoms}

We now wish to apply the diffractive periodic orbit theory in an 
atomic context by treating the non-hydrogenic core as a diffractive source.
The crucial step is to obtain an expression for the diffractive
constant $d(\theta_i,\theta_f)$.  

Consider a wave incident on the atomic core at an angle $\theta_f$ to
the $z$-axis.  On reaching the core, this wave produces a scattered
wave, $\psi_{\rm scatt}$, which feeds outgoing semiclassical waves
along periodic orbits;  this scattered wave can be decomposed  into 
two components:~\cite{GD92b}
\begin{equation}
\psi_{\rm scatt}(r,\theta) = 
\underbrace{\psi_{\rm Coul}(r,\theta \simeq \theta_i)}_{\rm Coulomb-scattered}
+ 
\underbrace{\psi^{\theta_f}_{\rm core}(r,\theta)}_{\rm Core-scattered}.
\end{equation}
The Coulomb-scattered wave, $\psi_{\rm Coul}$, is strongly
back-focussed and equated with the source for geometric paths.  
The core-scattered wave, $\psi_{\rm core}$, on the other hand, is
equated with the source of diffractive semiclassical waves.  At
some radius $r_0$, we express $\psi_{\rm core}$ in a partial wave 
expansion which, for $m=0$, is~\cite{GD92b} 
\begin{equation}
\psi_{\rm core}^{\theta_f}(r_0,\theta) = 
\left( \frac{2\pi^2}{r_0^3} \right )^{\frac14}
\sum_{l=0}^{\infty}  
Y_{l0}^{*}(\theta_f,0) Y_{l0}(\theta,0)( e^{2i\delta_l} - 1 ) 
e^{i(\sqrt{8r_0}-3\pi/4)} 
\label{theory.7}
\end{equation}
where $\delta_l$ indicates a set of $l$-dependent 
quantum defects describing the non-hydrogenic core.  
Finally, we take $d$ to be the 
fractional amplitude scattered by the core:
\begin{equation}
d(\theta_i,\theta_f)= \psi^{\theta_f}_{\rm core}(r_0,\theta_i)/
\psi_{\rm Coul}(r_0, \theta_f). 
\end{equation}

For small atoms only the lowest partial waves have 
non-zero quantum defects and hence contribute to the sum in
Eq.~(\ref{theory.7}). For example,  for even parity lithium, 
$\delta_{0} \simeq 0.4\pi$ and $\delta_{l \geq 2}\simeq 0$.
For such $s$-wave scattering, $\psi^{\theta_f}_{\rm core}$ is
isotropic.  All our calculations have been carried out for the case
of $s$-wave scattering so below $\delta\equiv\delta_{l=0}$;  
generalization to odd parity spectra and atoms with more than one 
non-zero quantum defect is straightforward.

For the case of $s$-wave scattering,
each diffractive contribution in Eq.~(\ref{diffrac}) is:
\begin{equation}
dG = 
\hbar^{1/2} (e^{2i\delta}-1) 
\left | \frac{2\pi}{m_{12}} 
\sin\frac{\theta_{i}}{2} \sin \frac{\theta_{f}}{2}\right | ^{1/2} 
e^{i\left (S/\hbar - \mu\pi/2 -\pi/4 \right ) }
\label{diff2}
\end{equation}
and, in effect, represents the contribution of a  pure diffractive orbit. 
Note the additional phase of $-\pi/4$
relative to an equivalent primitive geometric periodic 
orbit.  Eq.~(\protect{\ref{diff2}}) supersedes an 
expression given previously~\cite{pecam} which merely gave an estimate
of the fractional reduction in amplitude for $R_1$.

\begin{figure}[htb]
\centerline{
\psfig{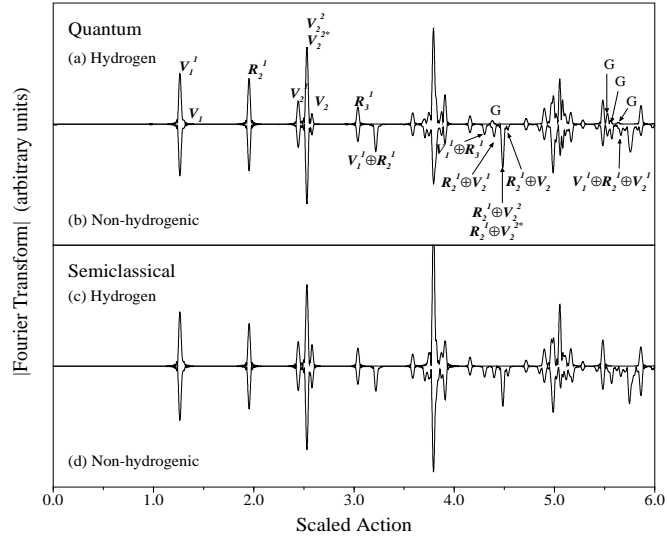}
}
\caption {(a) Comparison of Fourier transforms of the density of states 
for hydrogen and a non-hydrogenic atom with $\delta=0.5\pi$ in a 
static magnetic field
from a fully quantal calculation with average $\hbar=1/90$ at
$\varepsilon=-0.2$.  
(b) Comparison between quantal and semiclassical difference
spectra obtained by coherently subtracting the Fourier transforms
shown in (a).}
\label{Fig1}
\end{figure}
In Fig.~\ref{Fig1}(a) we show Fourier transforms of the oscillatory 
part of the even-$l$, $m=0$, eigenvalue spectra
for hydrogen and a non-hydrogenic atom with $\delta=0.5\pi$ in a static 
magnetic field at scaled energy
$\epsilon=-0.2$ and $n= \gamma^{-1/3}=\hbar^{-1}$ ranging from $60$ to
$120$, so the average value of $\hbar^{-1}=90$. In Fig.~\ref{Fig1}(b)
we plot the difference spectrum obtained by coherently subtracting the 
Fourier transform of the hydrogenic spectrum from that of the 
non-hydrogenic;  this exposes
the diffractive contributions and eliminates contributions from periodic 
orbits which do not pass through the core.  
For comparison, we also plot a semiclassical
difference spectrum obtained by summing all terms of order $\sqrt{\hbar}$ and
$\hbar$; agreement is excellent. The discrepancy in $D_2$
is due to the effects of bifurcations that are not taken into
account in the semiclassical calculation presented here.

We can see that for the non-hydrogenic case 
the amplitudes of the Garton-Tomkins orbit, $R_1$ and its harmonic $R_2$,
as well as the balloon, $V_1^1$, and other orbits are substantially reduced.
There are additional small peaks which correspond
accurately to sums of periodic orbits. Importantly, there are
strong peaks (marked $D_1$ and $D_2$) which do not match any combination
of orbits. At these scaled actions ($S\simeq 2.87$ and $S\simeq 2.94$) 
we find  orbits
that are closed but not periodic. For hydrogen, only orbits that are
periodic can contribute. Here we see
that pure diffractive orbits, such as $D_2$, can contribute to the
non-hydrogenic spectrum at
$O(\sqrt{\hbar})$ so are substantially stronger than combinations of 
orbits.  The peak at $S\simeq  2.87$ is due to an isolated closed orbit 
and is obtained almost exactly from Eq.~(\ref{diff2}) as seen in 
Fig.~\ref{Fig1}(b) (note that in Fig.~\ref{Fig1}(a) the peak associated 
with this  orbit is masked by the peak of a periodic orbit
which does not approach the nucleus). The peak at $S\simeq 2.94$
consists of contributions from a pair of non-isolated orbits close to a 
bifurcation so their contribution is over-estimated semiclassically. 
On examination of the diffractive orbits we find that they
correspond to the first closure of asymmetric periodic orbits,
some of which correspond to the $X_n$ series of `exotic orbits'.\cite{H88}
In hydrogenic eigenvalue spectra such orbits can only contribute at their
{\em full period,\/} whereas in the diffractive case they appear
at {\em closure.\/}

\section{Conclusions} 
We have now successfully extended our 1995 study of closed
orbit theory and for non-hydrogenic spectra to a study of periodic orbit theory
{\em per se,} in the form of the Gutzwiller trace formula.  We have
carried out a 
detailed study of diffraction in atoms in the GTF for several scaled
energies to study the $\hbar$ and $\delta$ dependence of the diffractive
effects.  We find that, away from bifurcation effects, our simple diffractive
corrections describe the quantal spectra to within an accuracy of about
$1\%$.  We note that a model-potential simulation with fully chaotic
dynamics gives agreement only to within a factor of about two in general.
We conclude that since the atomic core is of the order of one atomic unit,
the diffractive effect is too weak be simulated accurately by the
cumulative effect of numerous unstable orbits.  However, 
for low energy, it is most naturally and accurately described as 
a combination of stable motion, coupled with a breakdown of the usual
semiclassics due to diffraction.

\section*{Acknowledgements}

We thank Dominique Delande and Martin Gutzwiller for useful
discussions. We acknowledge support from the EPSRC.

\end{document}